\documentclass[11pt,twoside]{stmEPS}
\usepackage{amsbsy,amsmath,amssymb}

\newcommand{\doo}[2]{\ensuremath{{\partial #1\over \partial #2}}}
\newcommand{\be}{\begin{equation}}
\newcommand{\ee}{\end{equation}}
\usepackage{amsbsy,amsmath,amssymb,epsfig}
\newcommand{\fundoo}[2]{\ensuremath{{\delta #1\over \delta #2}}}

\def\vk{{\bf k}}

\def\vq{{\bf q}}

\def\vx{{\bf x}}

\begin{document}
\markboth{\sc J. Honkonen}{\sc Ito and Stratonovich calculuses in stochastic field theory}

\STM

\title{Ito and Stratonovich calculuses in stochastic field theory}

\authors{Juha~Honkonen}

\address{National Defence University, P.O.Box 7, FI-00861 Helsinki, Finland}
\bigskip

\begin{abstract}
Ambiguities in the functional-integral solution of the stochastic differential
equation (SDE) arising due to
the definition on the functional Jacobi determinant and the white-in-time limit in
the noise are analyzed and two forms of the de Dominicis-Janssen dynamic action proposed
corresponding to the Ito and Stratonovich interpretations of the SDE.
\end{abstract}

\section*{Introduction}
\label{sec:intro}

Since the invention of the functional-integral form \cite{DeDominicis76} of the Martin-Siggia-Rose approach to
solution of stochastic problems \cite{MSR} ambiguities in this formulation have been discussed
in the literature \cite{Munoz89,Zinn,Chaichian01,Vasilev04}. They have two sources: first, a change
of variables brings about a Jacobi determinant which is not well defined. Second,
the mathematical inconsistency of the Langevin equation with the white
noise leads to
different definitions of the SDE that show in the functional-integral solution.
The aim of this report is to analyze both cases in unified
manner. Two different ways to resolve the
ambiguities in the functional integral are proposed in the terms of dynamic actions in the Ito and Stratonovich forms.

\section*{Dynamic field theory with additive random field}
\label{sec:dynfieldth}

Consider the Langevin equation
\be
\label{Langevin}
\doo{\varphi}{t}=V(\varphi)+f:=-K\varphi+U(\varphi)+f\,,
\ee
where $f$ is a Gaussian random field with zero mean and the
white-in-time correlation function
\be
\label{correlator}
\langle f(t,{\bf x})f(t',{\bf
x}')\rangle=\overline{D}(x-x')=\delta(t-t')D({\bf x}-{\bf x}')\,,
\ee
where the shorthand notation $x=(t, {\bf x})$ has been used.

In (\ref{Langevin}) $K$ is usually a second-order differential operator in space
and $U({\bf x},\varphi)$ a nonlinear local in time functional of $\varphi$.
The paradigmatic example is model A of critical dynamics \cite{Hohenberg77}
defined by
\begin{equation}
\label{modelA}
\doo{\varphi}{t}=-\Gamma\left(-\nabla^2\varphi+a\varphi+{\lambda\over 6}\varphi^3\right)+f\,,
\end{equation}
and the correlation function $D({\bf x}-{\bf x}')=2\Gamma \delta({\bf x}-{\bf x}')$.

The Langevin equation with white-in-time noise $f$ is mathematically inconsistent, because
the time integral of the noise $\int f dt$ is a Wiener process which not differentiable anywhere.
To arrive at this limit, use the set of correlation functions
constituting a $\delta$ sequence in time, i.e.
\be
\label{deltaCorrAdditive}
\langle f(t,{\bf x})f(t',{\bf x}')\rangle=\overline{D}(t,{\bf x};t',{\bf x}')
\xrightarrow[t' \to t]{\hbox{}}\delta(t-t')D({\bf x},{\bf x}')\,.
\ee
This treatment gives rise to the solution of the SDE (\ref{Langevin}) in the Stratonovich sense \cite{Gardiner97}.

To cast the problem (\ref{Langevin}) and (\ref{deltaCorrAdditive}) into a functional-integral form,
use standard tricks:
\begin{multline}
G(J)=\langle e^{\varphi[f] J} \rangle=
\int\!{\cal D}\varphi\,\left\langle\delta\left(\varphi-\varphi[f]\right)\right\rangle\, e^{\varphi J}\\=
\int\!{\cal D}\varphi\,\left\langle\delta\left[-\partial_t\varphi+V(\varphi)+f\right]\vert\det M\vert\right\rangle
\, e^{\varphi J}\\
=\int\!{\cal D}\varphi\,\int\!{\cal D}\tilde{\varphi}\,\left\langle \,\vert\det M\vert\,
e^{\tilde{\varphi}\left[-\partial_t\varphi+V(\varphi)+f\right]}\right\rangle e^{\varphi J}\,.
\label{EqChain}
\end{multline}
Representation the Jacobi determinant in (\ref{EqChain})
in the loop expansion of $\displaystyle e^{{\rm Tr}\ln M}$ yields
\be
\label{DetMExp}
\det M =\det\left[\left(\frac{\partial }{\partial t}+K\right)\delta(x-x')-\frac{\delta
U({\bf x},\varphi)}{\delta\varphi(x')}\right]:=
\det\left(\partial_t+K\right)e^{-\Delta(0)U'}\,,
\ee
in which the diagonal value  $\Delta(0)$ of the response function of $\varphi$
is a parameter of the model. The response function is the retarded Green function of the free-field
equation, whose value at $t=t'$ is not determined.
The two popular choices are $\Delta(0)=0$~\cite{Vasilev04}
and $\Delta(0)={1\over 2}\delta({\bf x}-{\bf x}')$~\cite{Zinn}.
Contrary to claims in the literature \cite{Munoz89,Chaichian01}, the
determinant (\ref{DetMExp}) has nothing to do with the stochastic nature of the Langevin equation (\ref{Langevin}): it appears
in the functional-integral representation of the solution of the corresponding deterministic equation as well.

Thus, we arrive at the De Dominicis-Janssen action \be \label{DDJf}
S[\varphi,\tilde{\varphi},f]=-\Delta(0)U'+ \ln
P[f]+\tilde{\varphi}\left[-\partial_t\varphi-K\varphi+U(\varphi)+f\right]\,,
\ee where $P[f]$ is the probability density function of the field
$f$. The effect of the determinant (\ref{DetMExp}) is to cancel
superfluous graphs brought about by the usual Feynman rules for the
dynamic action (\ref{DDJf}). For instance, the iterative solution of
(\ref{modelA}) for the field $\varphi$ may be expressed as the
tree-graph expansion \be \label{IterField}
\varphi=\negthickspace\negthickspace\negthickspace
\raisebox{-4.0ex}{ \epsfysize=1.5truecm \epsffile{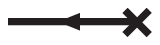}}+
\raisebox{-4.0ex}{ \epsfysize=1.5truecm \epsffile{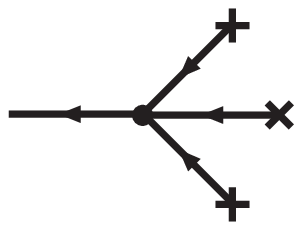}}+\ldots
\ee Here, the directed line corresponds to the propagator $\Delta$,
the cross to the field $f$ and the full dot to the vertex factor
\cite{Vasilev98} of the dynamic action (\ref{DDJf}). However,
according to the Feynman rules for (\ref{DDJf}), in the linear order
in $\lambda$, this solution should contain also a graph with a
closed propagator loop and graph brought about by the determinant
factor (here, the source field $f$ is fixed and $\ln P[f]=0$), i.e.
\be \label{PThField}
\varphi=\negthickspace\negthickspace\negthickspace
\raisebox{-4.0ex}{ \epsfysize=1.5truecm \epsffile{propagatorX.EPS}}+
\raisebox{-4.0ex}{ \epsfysize=1.5truecm
\epsffile{3tree.EPS}}+\negthickspace\negthickspace\negthickspace
\negthickspace\negthickspace\negthickspace\negthickspace\negthickspace\negthickspace
\raisebox{-4.0ex}{ \epsfysize=2.1truecm
\epsffile{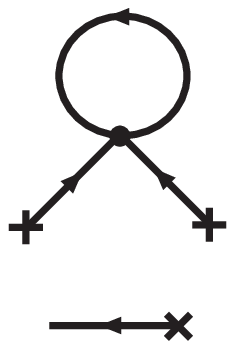}}\negthickspace\negthickspace\negthickspace
\negthickspace\negthickspace\negthickspace\negthickspace\negthickspace\negthickspace
+\negthickspace\negthickspace\negthickspace
\negthickspace\negthickspace\negthickspace\negthickspace\negthickspace\negthickspace
\raisebox{-4.0ex}{ \epsfysize=2.1truecm
\epsffile{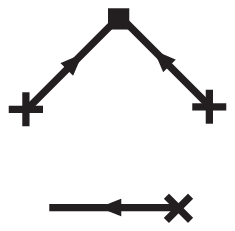}}\negthickspace\negthickspace\negthickspace
\negthickspace\negthickspace\negthickspace\negthickspace\negthickspace\negthickspace
+\ldots \ee The black square stands for the vertex factor brought
about by the determinant term
$-\Delta(0)U'=\Delta(0){1\over2}\lambda\Gamma\varphi^2$. The
coefficient $\Delta(0)$ is the value of the closed propagator loop
of the third graph in the right side of (\ref{PThField}), therefore
the third and fourth graphs in (\ref{PThField}) cancel each other
thus restoring the iterative solution (\ref{IterField}) at this
order.

Although the functional integral gives rise to the perturbative
solution independent of the value of $\Delta(0)$, the form of the dynamic
action is important for its nonperturbative calculation.
The choice
of $\det M$ affects the explicit form of the integrand of the functional integral in a nontrivial way.
The subsequent implicit dependence of the
functional integral on the parameter $\Delta(0)$ should be dealt with in nonperturbative approaches
such as numerical evaluation or instanton calculus \cite{Chaichian10,Honkonen05} to arrive at generating
function independent of $\Delta(0)$.
To be on the safe side, for purposes of any nonperturbative
calculation, the Jacobian should be taken in the
field-independent form
$
\det M =\det\left(\partial_t+K\right)\,,
$
which corresponds to the choice $\Delta(0)=0$.

\section*{Dynamic field theory with multiplicative random field}
\label{sec:multinoise}

A different picture emerges for a SDE with a multiplicative noise. Consider the simplest case
\be
\label{LangevinMultiLinear}
\doo{\varphi}{t}=-{K}\varphi +f\varphi\,,
\ee
where ${K}$ is a time-independent operator acting on the field $\varphi$ and
$f$ a Gaussian random field with zero mean. Difficulties in the interpretation of
the equation (\ref{LangevinMultiLinear}) arise, when the correlation function of the field $f$ is local
in time as in (\ref{correlator}), which is limiting case of the sequence
(\ref{deltaCorrAdditive}).

The iterative solution of the SDE (\ref{LangevinMultiLinear}) may
expressed as the series \be \label{solution}
\varphi=\Delta{\chi}+\Delta f \Delta{\chi} + \Delta f \Delta
f\Delta{\chi}+ \ldots \ee where ${\chi}$ is the initial condition of
the solution $ \Delta {\chi}=\int\! d{\bf x}'\Delta(t,{\bf x}-{\bf
x}'){\chi}({\bf x}') $ of the homogeneous equation
(\ref{LangevinMultiLinear}) where $\Delta$ is the (retarded) Green
function of the same equation. Graphically, the solution
(\ref{solution}) is a sum of chains of oriented lines corresponding
to retarded propagators \be \label{GraphSolution}
\varphi=\negthickspace\negthickspace\negthickspace
\raisebox{-4.0ex}{ \epsfysize=1.5truecm \epsffile{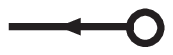}}+
\raisebox{-4.0ex}{ \epsfysize=1.5truecm \epsffile{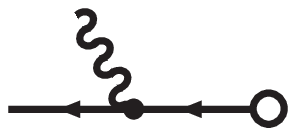}} +
\raisebox{-4.0ex}{ \epsfysize=1.5truecm \epsffile{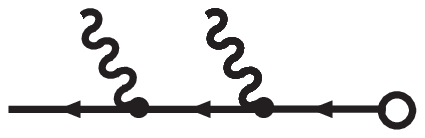}}+\
\ldots \ee where the circle stands for the initial condition
${\chi}$ of the homogeneous equation (\ref{LangevinMultiLinear}),
the wavy line corresponds to the random field $f$ and the full dot
represents the vertex factor brought about by the last term of
(\ref{LangevinMultiLinear}). The functional-integral representation
(\ref{EqChain}) for the solution of this problem gives rise to the
dynamic action \be \label{DDJfMulti} S[\varphi,\tilde{\varphi},f]=
-\Delta(0)f+\ln
P[f]+\tilde{\varphi}\left(-\partial_t\varphi-{K}\varphi
+f\varphi\right) \ee again with a term brought about by the
determinant factor $ \det M
=\det\left(\partial_t+K\right)\exp\left[-\Delta(0)f\right] $ which
has the same effect of cancelation of graphs as in the case of
additive external field. The perturbation-theory expression for the
field $\varphi$ at the order depicted in (\ref{GraphSolution}) is
\begin{multline}
\label{GraphsExcluded} \varphi=\negthickspace\negthickspace
\raisebox{-4.0ex}{ \epsfysize=1.5truecm \epsffile{emptydot.EPS}}+
\raisebox{-4.0ex}{ \epsfysize=1.5truecm \epsffile{onelinedot.EPS}} +
\raisebox{-4.0ex}{ \epsfysize=1.5truecm \epsffile{twolinedot.EPS}} +
\negthickspace\negthickspace \raisebox{-1.1ex}{ \epsfysize=1.5truecm
\epsffile{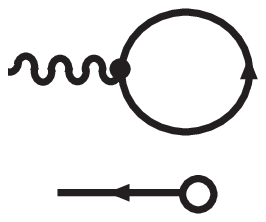}} + \raisebox{-1.0ex}{
\epsfysize=1.5truecm \epsffile{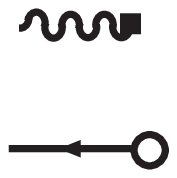}}
 +
\raisebox{-0.3ex}{ \epsfysize=1.5truecm
\epsffile{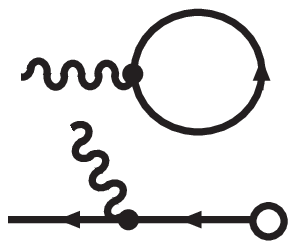}}\\
+ \raisebox{-0.3ex}{ \epsfysize=1.5truecm
\epsffile{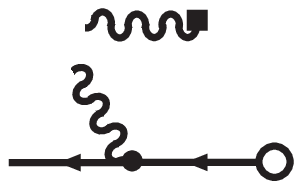}} \negthickspace + {1\over 2}
\raisebox{-.8ex}{ \epsfysize=1.5truecm
\epsffile{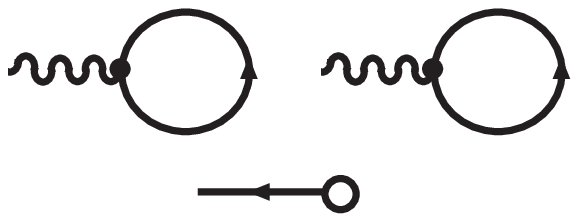}} \negthickspace\negthickspace
 + {1\over 2}
\raisebox{-.7ex}{ \epsfysize=1.5truecm
\epsffile{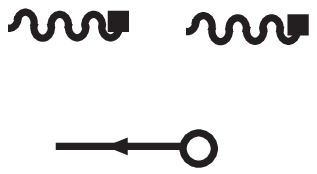}} + \raisebox{-.7ex}{
\epsfysize=1.5truecm \epsffile{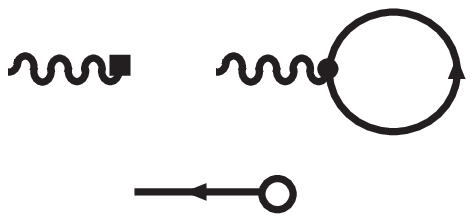}}
\negthickspace\negthickspace+ \,\ldots
\end{multline}
which again contains closed loops of the propagator.
Here, the black square stands for the $-\Delta(0)$ vertex factor.
Thus, the sum of the closed propagator loop graphs and the determinant-generated
graphs in the right side of (\ref{GraphsExcluded}) vanishes.
Note that the effect of the Jacobi
determinant to remove ''superfluous'' graphs with closed
propagator loops is independent of the nature of the source field $f$.

The average of the series (\ref{solution}) over the probability
density of $f$ gives the solution of the SDE (\ref{LangevinMultiLinear}) for the field $\varphi$.
For the Gaussian distribution with zero mean and the correlation function (\ref{correlator})
this gives rise to the dynamic action
\be
S[\varphi,\tilde{\varphi},f]=-\Delta(0)f -{1\over
2}fD^{-1}f+\tilde{\varphi}\left(-\partial_t\varphi-{K}\varphi
+f\varphi\right)\,.\nonumber \ee The analysis of the corresponding
functional integral is more transparent in terms of the three fields
$\varphi$, $\tilde{\varphi}$ and $f$. The perturbative solution of
the SDE (\ref{LangevinMultiLinear}) is given by Wick's theorem for
the Gaussian distribution of $f$, which graphically amounts to
replacing any pair of $f$ by the correlation function $\overline{D}$
depicted by an unoriented line in all possible ways. For instance,
\be \label{GraphSolutionAve}
\langle\varphi\rangle=\negthickspace\negthickspace\negthickspace
\raisebox{-4.0ex}{ \epsfysize=1.5truecm \epsffile{emptydot.EPS}}+
\raisebox{-4.0ex}{ \epsfysize=1.5truecm \epsffile{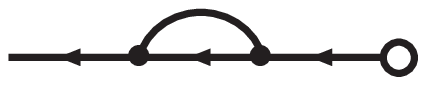}}+\
\ldots \ee In the limit of white-in-time correlations an enormous
truncation of the averaged iterative series (\ref{solution}) takes
place due to temporal $\delta$ functions contracting the ends of
chains of the retarded propagators. Only those terms, in which the
correlation function is multiplied by a single retarded propagator
do not vanish automatically.

For instance, the one-loop graph in (\ref{GraphSolutionAve}) in case of white-in-time noise
gives rise to an ambiguity, which is directly related
to that in the interpretation of the SDE (\ref{LangevinMultiLinear}).
A straightforward substitution of the white-noise correlation function in this graph gives rise to the expression
\begin{multline}
\raisebox{-4.0ex}{ \epsfysize=1.5truecm \epsffile{oneloopdot.EPS}} =
\int\!dt_1\int\!d{\bf x}_1\int\!d{\bf x}_2\int\!d{\bf x}_3\,\Delta(t-t_1,{\bf x}-{\bf x}_1)\\
\times
\Delta(0,{\bf x}_1-{\bf x}_2)D({\bf x}_1,{\bf x}_2)\Delta(t_1,{\bf x}_2-{\bf x}_3){\chi}({\bf x}_3)\,,\nonumber
\end{multline}
where the value of the propagator at coinciding
time arguments $\Delta(0,{\bf x}_1-{\bf x}_2)=\theta(0)\delta({\bf x}_1-{\bf x}_2)$ is again ambiguous,
although here the spatial $\delta$ function does not create a problem.

With the use of the $\delta$-sequence of correlation functions (\ref{deltaCorrAdditive}) this ambiguity
is readily resolved and gives rise to the expression
\begin{multline}
\label{1multiExStrato} \raisebox{-4.0ex}{ \epsfysize=1.5truecm
\epsffile{oneloopdot.EPS}} ={1\over 2} \int\!dt_1\int\!d{\bf
x}_1\int\!d{\bf x}_3\,\Delta(t-t_1,{\bf x}-{\bf x}_1) D({\bf
x}_1,{\bf x}_1)\Delta(t_1,{\bf x}_1-{\bf x}_3){\chi}({\bf x}_3)\,,
\end{multline}
with the coefficient ${1\over 2}$ in front of the spatial $\delta$ function.
Formally, this result may be obtained by amending the definition of the propagator
according to the rule $\Delta(0,{\bf x}-{\bf x}'))={1\over 2}\delta({\bf x}-{\bf x}')$. Then, of course, it would seem
appropriate to use the same rule in the Jacobi determinant factor. Within this choice the graphical
expression in (\ref{1multiExStrato}) appears excessive, because it hints to twice the number of integrations
than actually is carried out.

It is more convenient to adopt the convention
$\Delta(0)=0$ here as well. In this case the deterministic part of the dynamic action (\ref{DDJfMulti})
has to be amended to produce the correct nonvanishing result for graphs like (\ref{1multiExStrato}).
This may be effected by adding to the dynamic action the term
\be
\label{ActionAmendLinear}
\Delta S={1\over 2}\,\tilde{\varphi}D(0)\varphi:=
{1\over 2}\,\int\! dt\int\!d{\bf x}\,\tilde{\varphi}(t,{\bf x})D({\bf x},{\bf x})
{\varphi}(t,{\bf x})\,.
\ee
Graphically, this amounts to replacing the one-loop graph with the noise correlation function
by a new vertex factor with the coefficient ${1\over 2}D(0)$:
\[
\raisebox{-4.0ex}{ \epsfysize=1.5truecm
\epsffile{oneloopdot.EPS}}\to \raisebox{-4.0ex}{
\epsfysize=1.5truecm \epsffile{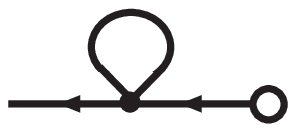}}\,.
\]
It may be readily seen that the amendment
(\ref{ActionAmendLinear}) to the dynamic action corresponds to the reorganization of the
SDE (\ref{LangevinMultiLinear}), when the Stratonovich equation is replaced by the Ito equation with the
same solution \cite{Gardiner97}. Thus, we arrive at the dynamic action in the Ito form
\be
\label{MultiIto}
S'[\varphi,\tilde{\varphi},f]=
-{1\over 2}fD^{-1}f+\tilde{\varphi}\left[-\partial_t\varphi-{K}\varphi +{1\over 2}\,D(0)\varphi+f\varphi\right]\,,
\quad \Delta(0)=0\,.
\ee
On the other hand, if we adopt the choice $\Delta(0,{\bf x}-{\bf x}')={1\over 2}\delta({\bf x}-{\bf x}')$, then it is natural to make the
same choice in the determinant factor and include it in the dynamic action. In this case,
the determinant term in the dynamic action leads to cancelation of terms containing
diagonal values of the propagator
$
\Delta(0)\int dt\int d{\bf x}\int d{\bf x}'D({\bf x}-{\bf x}')\Delta(0)={1\over 4}\,\delta^2(0)TV\int d{\bf x}D({\bf x})
$
but leaves intact terms containing a similar expression but with diagonal in time values of propagators contracted with different spatial arguments:
$
\int dt\int d{\bf x}\int d{\bf x}'\Delta(0,{\bf x}-{\bf x}')D({\bf x}-{\bf x}')\Delta(0,{\bf x}'-{\bf x})={1\over 4}\,\delta(0)TVD(0)\,.
$
To exclude the latter from perturbation theory,
the term
$
\Delta S'=-{1\over 8}\,\delta(0)T VD(0)
$
must be added to the determinant part of the dynamic action.
In this case we arrive at the
dynamic action in the Stratonovich form
\be
\label{MultiStrato}
S''[\varphi,\tilde{\varphi},f]=-{1\over 2}\,\delta(0)\left[{1\over 4}D(0)+f\right]
-{1\over 2}fD^{-1}f+\tilde{\varphi}\left[-\partial_t\varphi-{K}\varphi +f\varphi\right]\,.
\ee
In case of multiplicative noise the choice of the diagonal value
of the propagator affects the dynamic action in a fashion analogous to the relation between Ito and Stratonovich
SDE's. It should be emphasized that both dynamic actions (\ref{MultiIto}) and (\ref{MultiStrato})
give rise to the same correlation and response functions which correspond
to the Stratonovich solution of the SDE (\ref{LangevinMultiLinear}). With the use of similar
arguments, two dynamic actions may be constructed for the solution of
the SDE in the Ito from as well. Different dynamic actions for the functional-integral solution
of a SDE represent ambiguities in the construction of the functional integral, but not in the SDE itself.

The analogy between the compilation of dynamic actions and SDE
holds also in the case,
when the random source field is multiplied by a polynomial
function(al) of the field $\varphi$. Consider the SDE
\be
\doo{\varphi}{t}=-{K}\varphi +{1\over 2}f\varphi^2\nonumber
\ee
as an example. The dynamic action with the coloured noise is
\be
\label{DDJfMulti2}
S[\varphi,\tilde{\varphi},f]=-\Delta(0)f\varphi -{1\over 2}f\overline{D}^{\,\,-1}f+
\tilde{\varphi}\left(-\partial_t\varphi-{K}\varphi +{1\over
2}\,f\varphi^2\right)\,,
\ee
where the first term in the right side
is brought about by the determinant factor. Instead of
(\ref{GraphSolutionAve}) the first terms of the solution are (take
the $\delta$ sequence of the correlation functions of $f$)
\be
\label{PolyGraphSolutionAve}
\langle\varphi\rangle=\negthickspace\negthickspace\negthickspace
\raisebox{-4.0ex}{ \epsfysize=1.5truecm \epsffile{emptydot.EPS}}+
\raisebox{-4.0ex}{ \epsfysize=1.5truecm
\epsffile{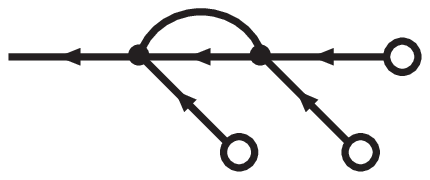}}+
\ \ldots
\ee
The analytic
expression of the second term in the white-noise limit assumes the form
\begin{multline}
\raisebox{-4.0ex}{ \epsfysize=1.5truecm \epsffile{oneloopdot2.EPS}}
={1\over 2}
\int\!dt_1\int\!d{\bf x}_1\int\!d{\bf x}_2\int\!d{\bf x}_3\int\!d{\bf x}_4\,\Delta(t-t_1,{\bf x}-{\bf x}_1)\\
\times
D({\bf x}_1,{\bf x}_1)
\Delta(t_1,{\bf x}_1-{\bf x}_2){\chi}({\bf x}_2)\Delta(t_1,{\bf x}_1-{\bf x}_3){\chi}({\bf x}_3)
\Delta(t_1,{\bf x}_1-{\bf x}_4){\chi}({\bf x}_4)\,.\nonumber
\end{multline}
The same result is brought about by the term
$
\Delta S={1\over 2}\,\tilde{\varphi}D(0)\varphi^3
$
added to the dynamic action (\ref{DDJfMulti2}). The form of this term corresponds to the
transition from the Stratonovich SDE to the Ito SDE. Graphically, the
one-loop white-noise subgraph shrinks to a dot:
\[
\raisebox{-4.0ex}{ \epsfysize=1.5truecm
\epsffile{oneloopdot2.EPS}}\to \raisebox{-4.0ex}{
\epsfysize=1.5truecm \epsffile{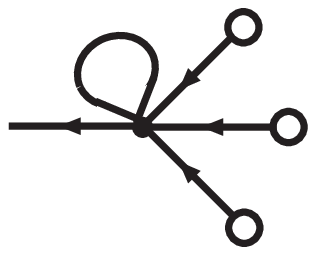}}\,.
\]
In case of Stratonovich convention,
terms with propagator loops contracted with different
spatial arguments are accounted for by the term
$
\Delta S'=-{1\over 8}\delta(0)D(0)\int dt\int d{\bf x}\varphi^2(x)
\,,
$
which should then be added to the dynamic action (\ref{DDJfMulti2}).

\section*{Iterative solution of the stochastic differential equation}
\label{sec:iterations}

Consider the SDE with a multiplicative white-in-time noise in the mathematically consistent form
$W$ of the form
\be
\label{SDE}
d\varphi=-K\varphi dt+b(\varphi)dW\,.
\ee
Here, $dW$ is the
increment of a Wiener process for each value of the spatial coordinate ${\bf x}$,
the deterministic part is linear in $\varphi$ for simplicity whereas the coefficient
function $b({\bf x},\varphi)$ of the stochastic part is a polynomial time-local functional of $\varphi$.

Let us construct a solution for the SDE (\ref{SDE}) by iterations. To this end,
introduce first spatial Fourier transform to obtain
\be
\label{SDEFourier}
\varphi(t,\vk)=\varphi(0,\vk)
-K\int\limits_0^t\!\varphi(t',\vk)dt'+
\int\limits_0^t\!\int\!{d\vq\over (2\pi)^d}\,b(\varphi,\vq)dW(t',\vk-\vq)\,,
\ee
where $K$ is now a constant and $b(\varphi,\vq)$ is the Fourier transform of the coefficient
function $b(\varphi)$ of (\ref{SDE}) and $dW(t,\vk)$ is the that of the increment of the Wiener process.
In (\ref{SDEFourier}) the function $b(\varphi)$ is expressed as a functional of the field $\varphi$ in the
Fourier representation. Construct then the iterative solution for the
Laplace transform $\phi(s,\vk)$
of the field $\varphi(t,\vk)$. The first two steps of the iteration process
yield
\[
\phi_0(s,\vk)={\varphi(0,\vk)\over s+K}\,,\quad
\phi_1(s,\vk)={1\over s+K}\,\int\limits_0^\infty\!e^{-st}
\int\!{d\vq\over (2\pi)^d}\,b(\phi_0,\vq)dW(t,\vk-\vq)\,.
\]
Here, $\varphi(0,\vk)$ is the Fourier transform of the initial value of the problem.
The iterative solution of (\ref{SDEFourier}) is readily expressed in terms of a tree-graph expansion of the type
(\ref{GraphSolution}), when the coefficient function $b(\varphi)$ is a polynomial function of $\varphi$.

The ambiguity in the solution of (\ref{SDE}) shows, when averages of functions of $\varphi$ are calculated.
The combinatorial rule for the Wiener process is given by the usual Wick theorem. The result of calculation
of time integrals of pairwise products of the Wiener process in case of Ito stochastic integral is
given by the correlation formula \cite{Gardiner97}
\be
\Biggl\langle\ \int\limits_{t_0}^t\!G(u)dW(u)\int\limits_{t_0}^t\!H(v)dW(v)\Biggr\rangle=
\int\limits_{t_0}^t\!\langle G(u)H(u)\rangle du\,,\nonumber
\ee
where $G(t)$ and $H(t)$ are continuous nonanticipating functions and the brackets denote average
over the probability distribution of the Wiener process. All functions of time emerging
in the iterative solution of the SDE (\ref{SDE}) are nonanticipating.

The practical rule of calculation may be stated as follows \cite{Gardiner97}: rewrite the increment of
the Wiener process as
$
dW(t)\to \xi(t)dt\,,
$
where $\xi(t)$ is $\delta$-correlated in time, i.e.
$
\langle \xi(t)\xi(t')\rangle=\delta(t-t')\,,
$
and calculate the averages in terms of $\xi$. With the use of this rule, integrals appear in which the
$\delta$ function of correlations of $\xi$ is integrated in such a way that one of the limits of integration
coincides with one of the arguments of the $\delta$ function. With the use of Ito stochastic integral, ambiguities
arising in these case are resolved as
\be
\label{ItoRule}
\int\limits_{t_1}^{t_2}\!f(t)\delta(t-t_1)=f(t_1)\,,\quad
\int\limits_{t_1}^{t_2}\!f(t)\delta(t-t_2)=0\,.
\ee
In case of the Stratonovich stochastic integral these rules are replaced by
\be
\label{StratoRule}
\int\limits_{t_1}^{t_2}\!f(t)\delta(t-t_1)={1\over 2}f(t_1)\,,\quad
\int\limits_{t_1}^{t_2}\!f(t)\delta(t-t_2)={1\over 2}f(t_2)\,.
\ee
In the tree-graph iterative solution these situations arise solely
due to the presence of the temporal step function in the propagator. The Ito rules (\ref{ItoRule})
exactly correspond to the choice $\theta(0)=0$, whereas the Stratonovich rules (\ref{StratoRule}) correspond to the
choice $\theta(0)={1\over 2}$ for the value of the step function at the origin.

Thus, we see that averages of the iterative solution of the SDE (\ref{SDE}) in the Stratonovich sense
give rise to the same
expression as the perturbation expansion on the basis of the dynamic action
in the Stratonovich form:
\be
S''[\varphi,\tilde{\varphi},f]=-{1\over 8}\overbrace{b'Db'}-{1\over 2}\,b'f
-{1\over 2}fD^{-1}f+\tilde{\varphi}\left[-\partial_t\varphi-{K}\varphi +fb(\varphi)\right]\,,
\quad \Delta(0,{\bf x}-{\bf x}')={1\over 2}\delta({\bf x}-{\bf x}')\,.\nonumber
\ee
Here, the determinant contribution in an expanded form is
\be
{1\over 8}\overbrace{b'Db'}+
{1\over 2}\,b'f={1\over 8}\int\! dt\int\!d\vx\int\!d\vx'\fundoo{b(\vx,\varphi)}{\varphi(t,\vx')}D(\vx-\vx')\fundoo{b(\vx',\varphi)}{\varphi(t,\vx)}+
{1\over 2}\int\! dt\int\!d\vx \fundoo{b(\vx,\varphi)}{\varphi(t,\vx)}f(t,\vx)\,.\nonumber
\ee
It has been shown above that the same solution may be obtained in
the perturbation expansion based on the dynamic action in the Ito
form. A generalization of that argument to the case of arbitrary
monomial shows that for the case of
polynomial $b(\varphi)$ the additional term needed to obtain the Ito
form of the dynamic action corresponding to the SDE (\ref{SDE}) is
\be
\label{ExpandedItoTerm}
\Delta S={1\over 2}\,\tilde{\varphi}b'(\varphi)Db(\varphi)
={1\over 2}\int\! dt\int\!d\vx \int\!d\vx'\tilde{\varphi}(t,\vx)\fundoo{b(\vx,\varphi)}{\varphi(t,\vx')}D(\vx-\vx')
b(\vx',\varphi)\,.
\ee
Here, $D(\vx-\vx')$ is the correlation
function of the Wiener processes as functions of the coordinate,
i.e. $ \left\langle W(t,\vx)W(t,\vx')\right\rangle=tD(\vx-\vx') $
instead of the relation $\left\langle W(t)W(t)\right\rangle=t$ for a
single Wiener process.

Thus, the iterative solution of the SDE (\ref{SDE}) in the Stratonovich sense may be constructed
with use of the Ito dynamic action of the form
\be
S'[\varphi,\tilde{\varphi},f]=
-{1\over 2}fD^{-1}f+\tilde{\varphi}\left[-\partial_t\varphi-{K}\varphi
+{1\over 2}\,b'(\varphi)Db(\varphi)
+fb(\varphi)\right]\,,
\quad \Delta(0,{\bf x}-{\bf x}')=0\,,\nonumber
\ee
where we have used the condensed notation (\ref{ExpandedItoTerm}).

\section*{Dynamic action in the Ito form and the Stratonovich form}
\label{sec:actions}

The analysis of preceding sections allows to write down the following prescriptions for the
construction of the solution of the following Stratonovich SDE
\be
\label{StratoSDE}
\partial_t\varphi=-{K}\varphi+U(\varphi) +f\,b(\varphi)\,,
\ee
in which the deterministic part is written in a generic form with the
nonlinear term $U(\varphi)$.

The solution of (\ref{StratoSDE}) is brought about by the functional-integral representation
either with the dynamic action in the Ito form with $\Delta(0,{\bf x}-{\bf x}')=0$
\be
S'[\varphi,\tilde{\varphi},f]=
-{1\over 2}fD^{-1}f
+\tilde{\varphi}\left[-\partial_t\varphi-{K}\varphi+U(\varphi)
+{1\over 2}\,b'(\varphi)Db(\varphi)
+fb(\varphi)\right]\,,\nonumber
\ee
or with the dynamic action in the Stratonovich form with $\Delta(0,{\bf x}-{\bf x}')={1\over 2}\delta({\bf x}-{\bf x}')$
\be
S''[\varphi,\tilde{\varphi},f]=
-{1\over 2}\left[U'(\varphi)+{1\over 4}\overbrace{b'Db'}+b'(\varphi)f\right]
-{1\over 2}fD^{-1}f
+\tilde{\varphi}\left[-\partial_t\varphi-{K}\varphi +U(\varphi)+fb(\varphi)\right]\,.\nonumber
\ee
From these representations the Ito is both more convenient in practical calculations in perturbation theory
and more reliable for use in non-perturbative evaluation of the functional integral.

Analogous representations for the action functional in stochastic dynamics defined by the Fokker-Planck equation
have been obtained also
within the operator approach of quantum field theory \cite{Leschke77} and interpolation construction
of the path integral \cite{Janssen}.

\section*{Conclusion}
\label{sec:conclusion}

Ambiguities appearing in the functional-integral solution of stochastic differential equations
have been analyzed. The representation of the generating function of the differential equation
as a functional integral gives rise to an ambiguity, which is completely independent of the statistics
of the source field: this ambiguity in the choice of the initial value of the propagator appears even in the
case of deterministic equation. The other ambiguity is due to the mathematical inconsistency of the SDE with
white-in-time noise, whose resolution gives rise to interpretations in the Ito sense and in the Stratonovich
sense. It has been shown that the functional-integral solution of both
SDEs may be constructed with the aid of two dynamic actions, which
differ by the choice of the
initial value of the propagator and the deterministic part of the dynamic action.
The difference of the deterministic parts of the two dynamic actions corresponds
to the connection between Ito and Stratonovich
SDEs having the same solution.

\section*{Acknowledgement}

I am indebted to Prof. H.-K. Janssen for bringing Refs. \cite{Leschke77,Janssen} to my attention.


\begin{thebibliography}{99}
\bibitem{DeDominicis76}
C. De Dominicis, J. Phys. (Paris) {\bf 37}, Suppl C1, 247 (1976);\\
H.K. Janssen, Z. Phys. B {\bf 23}, 377 (1976).
\bibitem{MSR}
P.C. Martin, E.D. Siggia, H.A. Rose, Phys. Rev. A {\bf 8}, 423 (1973).
\bibitem{Munoz89}
G. Mu\~{n}oz, W. S. Burgett, J. Stat. Phys. {\bf 56}, 59 (1989).
\bibitem{Zinn} J. Zinn-Justin, Quantum Field Theory and
Critical Phenomena, 3rd edition, Clarendon Press, Oxford, 1996.
\bibitem{Chaichian01}
M. Chaichian, A. Demichev,
Path Integrals in Physics,
Volume I,
Stochastic Processes and Quantum Mechanics, IOP Publishing, Bristol, 2001.
\bibitem{Vasilev04}
A.N. Vasil'ev, {The Field Theoretic Renormalization
Group in Critical Behavior Theory and Stochastic Dynamics}, Chapman \& Hall/CRC, Boca Raton, 2004.
\bibitem{Hohenberg77}
P.C. Hohenberg, B.I. Halperin, Rev. Mod. Phys. {\bf 49}, 435 (1977).
\bibitem{Gardiner97}
C.W. Gardiner, Handbook of Stochastic Methods for Physics, Chemistry and the Natural Sciences,
Springer, Berlin, 1997.
\bibitem{Vasilev98}
A.N. Vasiliev, Functional Methods in Quantum Field Theory and Statistical Physics,
Gordon and Breach, Amsterdam, 1998.
\bibitem{Chaichian10}
M. Chaichian, A. Tureanu, A. Zahabi,
Phys. Rev. E {\bf 81}, 066309 (2010).
\bibitem{Honkonen05}
J. Honkonen, M. V. Komarova, M. Yu. Nalimov,
Nucl. Phys. B {\bf 707} [FS], 493 (2005).
\bibitem{Leschke77}
H. Leschke, M. Schmutz, Z. Phys. B {\bf 27}, 85 
(1977).
\bibitem{Janssen}
H.-K. Janssen, On the Renormalized Field Theory of Nonlinear Critical Relaxation. In: From Phase Transitions to Chaos, Eds.
G. Gy\"orgyi, I. Kondor, L. Sasv\'ari, T. T\'el, World Scientific, London, 1991.
pp. 68-91.


\end{thebibliography}
\end{document}